\documentclass[12pt,final,fleqn]{article}

\usepackage[margin=1in] { geometry }
\usepackage{amssymb,amsmath, bm}
\usepackage{verbatim}
\usepackage[utf8]{inputenc}
\usepackage{setspace}
\usepackage{enumitem}
\usepackage[bottom]{footmisc}
\usepackage{url}
\usepackage[font={bf}]{caption}
\usepackage{float}

\usepackage{caption}
\usepackage{subcaption}
\usepackage{booktabs}
\usepackage[export]{adjustbox}

\usepackage{setspace}
\usepackage{latexsym}
\usepackage{graphicx}
\usepackage{marvosym}
\usepackage{amsmath} 
\usepackage{authblk}
\usepackage{xcolor}
\usepackage{blindtext}
\usepackage{threeparttable}
\usepackage{pdflscape}
\usepackage{tabularx}

\usepackage[american]{babel}
\usepackage[babel]{csquotes}
\usepackage[authordate,natbib,giveninits,backend=biber]{biblatex-chicago} 
\usepackage[toc,page,header]{appendix}

\usepackage{chngcntr}
\usepackage{etoolbox}
\usepackage{lipsum}
\usepackage{etoc}

\usepackage{graphicx}
\usepackage{rotating}
\usepackage{amsmath}

\usepackage{color}
\usepackage{hyperref}
\usepackage{xcolor}
\hypersetup{
    colorlinks,
    linkcolor={blue!50!black},
    citecolor={blue!50!black},
    urlcolor={blue!80!black}
}

\newcommand{\aref}[1]{\hyperref[#1]{Appendix~\ref{#1}}}

\title{\textbf{Dyadic Clustering in International Relations}}
\author{Jacob Carlson\thanks{Harvard University, Department of Economics. jacob\_carlson@g.harvard.edu.}, Trevor Incerti\thanks{University of Amsterdam, Department of Political Science. t.n.incerti@uva.nl.}, and P.M. Aronow\thanks{Yale University, Departments of Political Science, Statistics \& Data Science, Biostatistics, and Economics. p.aronow@yale.edu.}}
\date{\today}

\begin{document}

\maketitle

\begin{center}
{\large Forthcoming, \textit{Political Analysis}}
\end{center}
\vspace{1cm}

\begin{abstract}
\noindent
Quantitative empirical inquiry in international relations often relies on dyadic data. Standard analytic techniques do not account for the fact that dyads are not generally independent of one another. That is, when dyads share a constituent member (e.g., a common country), they may be statistically dependent, or ``clustered.'' Recent work has developed dyadic clustering robust standard errors (DCRSEs) that account for this dependence. Using these DCRSEs, we reanalyzed all empirical articles published in \textit{International Organization} between January 2014 and January 2020 that feature dyadic data. We find that published standard errors for key explanatory variables are, on average, approximately half as large as DCRSEs, suggesting that dyadic clustering is leading researchers to severely underestimate uncertainty. However, most (67\% of) statistically significant findings remain statistically significant when using DCRSEs. We conclude that accounting for dyadic clustering is both important and feasible, and offer software in \textsf{R} and Stata to facilitate use of DCRSEs in future research. 
\end{abstract}

\newpage

\section{Introduction}

Analysis of dyadic data---data in which each observation represents a dyad, or pair, of units, e.g., countries---is common in quantitative empirical research in international relations. Seminal theories such as the democratic trade hypothesis \citep{dixon1993political, bliss1998democratic, green2001dirty, mansfield2000free}, democratic peace theory \citep{oneal2001clear, dafoe2011statistical, imai2021robustness, gartzke2007capitalist}, liberal peace theory \citep{oneal2001clear}, and democratic alliance formation \citep{simon1996political, gibler2006alliances} claim empirical support (or lack thereof) from the analysis of dyadic data.

Dyadic data have a unique dependency structure---one where repeated observations of dyads are likely correlated with one another (as in panel datasets) \textit{and} dyads that share a common member are likely correlated with one another. In other words, because multiple dyads can share members, model errors can be correlated across dyads. Only accounting for the correlations between repeated observations of dyads (e.g., by using dyad clustered standard errors or fixed effects) and ignoring correlations across dyads assumes dyad-level events are independent. This assumption contradicts substantive knowledge of the dependencies between many types of dyads common to the social sciences, such as dyads of countries. As \citet{erikson2014dyadic} note, ``when a nation undergoes a pro-democratic revolution or, alternatively, when democratic leaders are deposed in a coup, the change ripples through all the nation’s many dyads.'' 

The idea that dyadic data exhibits a unique clustering structure that needs to be addressed methodologically in empirical work is not novel to political scientists. Random effects models have been proposed for dyads \citep{cameron2005estimation}, \citet{erikson2014dyadic} proposed a permutation testing framework that accounts for dyadic structure, and fully parametric analyses have accounted for dyadic structure and network structure \citep{hays2010spatial}. Previous research has therefore determined that failing to properly account for dyadic clustering may result in underestimation of the size of standard errors and confidence intervals \citep{aronow2015cluster, erikson2014dyadic, cameron2014robust}. However, these methodological insights have not yielded a corresponding change in the way in which applied scholars conduct their research. 

Recent work has developed standard error estimators that account for dyadic clustering. Building from \citet{fafchamps2007formation}, \citet{cameron2014robust}, \citet{aronow2015cluster}, and \citet{tabord2019inference} have developed and studied dyadic cluster-robust standard errors (DCRSEs). Using these DCRSEs, we reanalyzed all articles published in \textit{International Organization} over the course of just over six years (January 2014 – January 2020) that feature dyadic data, none of which originally implemented DCRSEs in their primary dyadic analyses. We find that DCRSEs are on average approximately twice as large as published standard errors, but that most findings remain statistically significant. While the literature therefore dramatically understates uncertainty, the estimated coefficients are usually large enough to remain statistically significant at conventional levels. To facilitate accounting for dyadic clustering in future research, we also offer software in both \textsf{R} and Stata that perform calculation of DCRSEs. 

Our primary contributions are therefore: (1) to empirically assess the degree to which uncertainty has been underestimated in previous research due to the presence of dyadic clustering, and (2) to increase the accessibility of potential solutions to dyadic clustering for applied scholars.

Note, however, that DCRSEs are not a panacea—the underlying theory and data generating process of the empirical setting should be taken into account prior to choosing an estimation strategy. When there is dependence between non-incident dyads (i.e., dyads that do not share a common member), DCRSEs will still underestimate uncertainty, just as only clustering on repeated dyads underestimates uncertainty when there is dependence across incident dyads. DCSREs should therefore not be considered a replacement for  approaches to clustering that (appropriately) account for greater amounts of dependence in the data.  The reanalysis we present therefore offers a lower bound on the severity of the consequences of inadequate clustering practices in previous research, or, in other words, reveals the extent to which dyadic clustering \textit{alone} is jeopardizing the reliability of research.

% well-motivated model-based approaches to clustering or other

\section{Why Is Dyadic Clustering a Problem?}

Dyadic data contain a dependency structure whereby repeated observations of dyads are allowed to be correlated with one another, and, importantly, dyads that share any common member are \textit{also} allowed to be correlated with one another. 

It may be helpful to illustrate the substantive assumptions implicit in assuming that dyad-level events are independent with common examples from IR theory. As \citet{cranmer2016critique} and \citet{maoz2019dyadic} note, assuming dyadic independence in WWII-era conflict implies that the conflict between Germany and Poland is unrelated to the conflict between Germany and Great Britain. This assumption is not realistic, as we know that Great Britain used the German invasion of Poland to justify its declaration of war on Germany. Similarly, \citet{neumayer2010spatial} and \citet{poast2016dyads} note that bilateral trade or investment treaties are influenced by the other treaties each member may already be a part of. Assuming independence would, for example, imply that a bilateral trade deal between the US and UK is unrelated to post-Brexit UK-EU trade negotiations. We provide an illustration using bilateral trade flows in the following section. Note that we set aside separate issues relating to analysis of time series cross sectional data \citep{beck1995and, blackwell2018make}.

\subsection{Toy Example: Bilateral Trade}

Consider an example in which we observe trade volume for a set of country-country-year dyads. For US-UK trade volume, any observations that include either the US or UK may be correlated with observations of any other dyad that also includes either the US or UK, respectively. \autoref{fig:clustering} illustrates the difference in assumptions about the dependencies between countries under traditional clustering by repeated dyad only, and with full dyadic clustering. \autoref{fig:clustering} highlights that under clustering by repeated dyad only, all country groups that do not share \textit{both} members are assumed to be independent. By contrast, under dyadic clustering, only country groups that share no members are assumed to be independent. 

\begin{table}[H]
\footnotesize
\centering
\begin{subtable}[t]{\linewidth}
\centering
\vspace{0pt}
\begin{tabular}{@{}lcccc@{}}
\toprule
\multicolumn{1}{c}{Dyad} & US-UK 1                            & US-UK 2                            & US-FR 1                            & UK-ES 1                            \\ \midrule
US-UK 1                  & Dependent                          & Dependent                          & {\color[HTML]{9B9B9B} Independent} & {\color[HTML]{9B9B9B} Independent} \\
US-UK 2                  & Dependent                          & Dependent                          & {\color[HTML]{9B9B9B} Independent} & {\color[HTML]{9B9B9B} Independent} \\
US-FR 1                  & {\color[HTML]{9B9B9B} Independent} & {\color[HTML]{9B9B9B} Independent} & Dependent                          & {\color[HTML]{9B9B9B} Independent} \\
UK-ES 1                  & {\color[HTML]{9B9B9B} Independent} & {\color[HTML]{9B9B9B} Independent} & {\color[HTML]{9B9B9B} Independent} & Dependent                          \\ \bottomrule
\end{tabular}

\caption{Clustering by Repeated Dyad Only}\label{tbl:repeated_dyad}

\vspace{0.25cm}

\begin{tabular}{@{}lcccc@{}}
\toprule
\multicolumn{1}{c}{Dyad} & US-UK 1   & US-UK 2   & US-FR 1   & UK-ES 1   \\ \midrule
US-UK 1                  & Dependent & Dependent & Dependent & Dependent \\
US-UK 2                  & Dependent & Dependent & Dependent & Dependent \\
US-FR 1                  & Dependent & Dependent & Dependent & {\color[HTML]{9B9B9B} Independent} \\
UK-ES 1                  & Dependent & Dependent &  {\color[HTML]{9B9B9B} Independent} & Dependent \\ \bottomrule
\end{tabular}
\caption{Dyadic Clustering}\label{tbl:dyadic}
\end{subtable}\hfill

\caption{Assumed Dependence by Clustering Type}\label{fig:clustering}
\end{table}

This clustering structure affects statistical inference. To illustrate this, suppose we were interested in characterizing the variance of the average level of commerce between the US and UK $(Y_{US-UK})$ and between the US and France $(Y_{US-FR})$. We can compute the sample mean of their outcomes: $\hat \mu = \frac{1}{2} (Y_{US-UK} + Y_{US-FR})$. \textit{If} we were to assume that dyads were statistically independent of one another, the variances are simply additive (i.e., there is no covariance term) and we could compute the variance of $\hat \mu$ as
\begin{align*}
\text{V}[\hat \mu]_{naive} &=  \text{V}\bigg[\frac{1}{2} (Y_{US-UK} + Y_{US-FR})\bigg] \\
                 &= \frac{1}{4} (\text{V}[ Y_{US-UK} ] + \text{V}[  Y_{US-FR} ]) .
\end{align*}

However, with dyadic data this may not be reasonable, especially when country-level factors are likely to impact dyadic outcomes. To see this, suppose that the true data generating process is additive among countries, so that

\begin{align*}
Y_{US,UK} &= aX_{US} + bX_{UK} + U_{US,UK} \\
Y_{US,FR} &= cX_{US} + dX_{FR} + U_{US,FR}
\end{align*}

\noindent
where the $X$ and $U$ are independent, and all $X$ are pairwise independent. In this instance, the true variance of $\hat \mu$ is

\begin{align*}
\text{V}[\hat \mu] &= \text{V}\bigg[\frac{1}{2} (Y_{US-UK} + Y_{US-FR})\bigg] \\
                  &=  \frac{1}{4} (\text{V}[ Y_{US-UK} ] + \text{V}[  Y_{US-FR} ]) + \frac{1}{2} \text{Cov}[ Y_{US-UK}, Y_{US-FR}] \\
	          &=    \text{V}_{naive} + \frac{1}{2} \text{Cov}[aX_{US} + bX_{UK} + U_{US,UK}, cX_{US} + dX_{FR} + U_{US,FR}] \\
	          &= \text{V}_{naive} + \frac{ac}{2} \text{V}[X_{US}] .
\end{align*}
When $a$ and $c$ are of the same sign---i.e., the dyads share a positive correlation---the naive characterization of the variance understates the true sampling variability. Our setting did not involve any “network effects” between countries: the problem emerges solely from the fact that a single country (here, the US) mechanically has more than one dyad in the data. 

\section{Standard Error Estimation with Dyadic Data}

Our approach to standard error estimation largely follows the logic of  \cite{cameron2014robust}, in which errors are likely correlated between dyad observations that have a country in common. To ease exposition, we consider the linear model of $Y_{ijt}$ on regressors $x_{ijt}$,
\begin{align*}
Y_{ijt} = \beta x_{ijt}  + u_{ijt}
\end{align*}

\noindent
where $Y_{ijt}$ is the level of commerce between countries $i$ and $j$ in time period or observation $t$ and $\beta$ is the slope that we obtain from fitting this model to the entire population. Under the exogeneity condition $\text{E}[u_{ijt} \mid x_{ijt}] = 0$ and the usual regularity conditions, the parameters of this model can be estimated using ordinary least squares.

The question is then how to estimate uncertainty in this model. Generically, the variance of an estimated parameter from a linear model can be represented in a symmetric form that resembles a sandwich, with two identical “bread” matrices and a “meat” matrix multiplied together in the order of bread, meat, and bread again \citep{greene2002econometric,  aronow2019foundations, davidson2004econometric}:

\begin{align*}
{V}_{sandwich} = (X^{T} X)^{-1} X^{T} \Omega X (X^{T} X)^{-1}
\end{align*}

\noindent
where $X$ denotes the matrix of regressors, and $\Omega$ is the variance-covariance matrix of model errors, such that $\Omega_{ijt,i’j’t’} = \text{E}[u_{ijt} u_{i’j’t’}]$. Robust sandwich estimators are formed by assuming that some elements of $\Omega$ are equal to zero, and then substituting residuals for errors, and means for expectations. Thus, the empirical variance-covariance matrix of model residuals, $\hat{\Omega}$, can be plugged into the above expression of variance to arrive at a variance estimator of the following form:\footnote{Both the Eicker-Huber-White heteroskedasticity-consistent variance estimator and the Liang and Zeger cluster robust variance estimator are sandwich-like, ``plug-in" variance estimators of this form, each making differing assumptions about $\Omega$.}

\begin{align*}
\hat{V} = (X^{T} X)^{-1} X^{T} \hat{\Omega} X (X^{T} X)^{-1} .
\end{align*}

\noindent
So long as there is not ``too much" clustering (for a precise statement, see \citealt{aronow2018confidence}), $\hat{V}$ will be a consistent estimator of the variance-covariance matrix of the sampling distribution of $\hat{\beta}$. The question then becomes, what restrictions on $\Omega$ are suitable for the problem at hand? The simplest case is to assume that there is no dependence across observations whatsoever in the data, which is the assumption for non-clustered RSEs: if $i \neq i’$, $j \neq j’$, \textbf{or} $t \neq t’$, then $\text{E}[u_{ijt} u_{i’j’t’}] = 0$, or that the errors are uncorrelated. With 6 observations, \autoref{tab: naive_approach} demonstrates the variance-covariance structure assumed by the naive approach:

\begin{table}[H]
\centering
\caption{Naive Approach (No Clustering)}
\label{tab: naive_approach}
\begin{tabular}{lllllll}
\hline
$ijt$ / $i'j't'$ & 111 & 112 & 211 & 212 & 221 & 222 \\
\hline
111 & $\text{E}[u_{111}^2]$   & 0   & 0   & 0   & 0   & 0   \\
112 & 0   & $\text{E}[u_{112}^2]$   & 0   & 0   & 0   & 0   \\
211 & 0   & 0   & $\text{E}[u_{211}^2]$   & 0   & 0   & 0   \\
212 & 0   & 0   & 0   & $\text{E}[u_{212}^2]$   & 0   & 0   \\
221 & 0   & 0   & 0   & 0   & $\text{E}[u_{221}^2]$   & 0   \\
222 & 0   & 0   & 0   & 0   & 0   & $\text{E}[u_{222}^2]$ \\
\hline
\end{tabular}
\end{table}

In practice, the naive approach is widely recognized as inappropriate in the context of international relations. It is expected that, for example, changes in bilateral trade relations will have impacts beyond the immediate dyad and across time periods. \\

The most common approach is clustering by dyad, where it is assumed that errors are correlated  when $i = i'$ and $j = j'$, regardless of time period. With the same 6 observations, \autoref{tab: clust_by_dyad} demonstrates the additional clustering permitted: 

\begin{table}[H]
\caption{Clustering by Repeated Dyad}
\label{tab: clust_by_dyad}
\centering
\begin{tabular}{lllllll}
\hline
$ijt$ / $i'j't'$ & 111 & 112 & 211 & 212 & 221 & 222 \\
\hline
111 & $\text{E}[u_{111}^2]$   & $\text{E}[u_{111} u_{112}]$   & 0   & 0   & 0   & 0   \\
112 & $\text{E}[u_{112} u_{111}]$   & $\text{E}[u_{112}^2]$   & 0   & 0   & 0   & 0   \\
211 & 0   & 0   & $\text{E}[u_{211}^2]$   & $\text{E}[u_{211} u_{212}]$   & 0   & 0   \\
212 & 0   & 0   & $\text{E}[u_{212} u_{211}]$   & $\text{E}[u_{212}^2]$   & 0   & 0   \\
221 & 0   & 0   & 0   & 0   & $\text{E}[u_{221}^2]$   & $\text{E}[u_{221} u_{222}]$   \\
222 & 0   & 0   & 0   & 0   & $\text{E}[u_{222} u_{221}]$   & $\text{E}[u_{222}^2]$ \\
\hline
\end{tabular}
\end{table}
\noindent In \autoref{tab: clust_by_dyad} above, we can see that while this approach does account for within-dyad correlations across time, all observations in the matrix of model errors that do not share both members of the dyad are still assumed to be uncorrelated (i.e., $\text{E}[u_{ijt} u_{i’j’t’}] = 0$).

By contrast, our approach only assumes independence when country pairs share no members. With the same 6 observations, we can see the clustering permitted by the dyadic clustering approach:
\begin{table}[H]
\caption{Dyadic Clustering}
\label{tab: dyadic_clust}
\centering
\begin{tabular}{lllllll}
\hline
$ijt$ / $i'j't'$ & 111 & 112 & 211 & 212 & 221 & 222 \\
\hline
111 & $\text{E}[u_{111}^2]$   & $\text{E}[u_{111} u_{112}]$   & $\text{E}[u_{111} u_{211}]$   & $\text{E}[u_{111} u_{212}]$   & 0   & 0   \\
112 & $\text{E}[u_{112} u_{111}]$   & $\text{E}[u_{112}^2]$   & $\text{E}[u_{112} u_{211}]$   & $\text{E}[u_{112} u_{212}]$   & 0   & 0   \\
211 & $\text{E}[u_{211} u_{111}]$  & $\text{E}[u_{211} u_{112}]$ & $\text{E}[u_{211}^2]$   & $\text{E}[u_{211} u_{212}]$   & $\text{E}[u_{211} u_{221}]$   & $\text{E}[u_{211} u_{222}]$   \\
212 & $\text{E}[u_{212} u_{111}]$   & $\text{E}[u_{212} u_{112}]$   & $\text{E}[u_{212} u_{211}]$   & $\text{E}[u_{212}^2]$   & $\text{E}[u_{212} u_{221}]$   & $\text{E}[u_{212} u_{222}]$   \\
221 & 0   & 0   & $\text{E}[u_{221} u_{211}]$   & $\text{E}[u_{221} u_{212}]$   & $\text{E}[u_{221}^2]$   & $\text{E}[u_{221} u_{222}]$   \\
222 & 0   & 0   & $\text{E}[u_{222} u_{211}]$   & $\text{E}[u_{222} u_{212}]$   & $\text{E}[u_{222} u_{221}]$   & $\text{E}[u_{222}^2]$ \\
\hline
\end{tabular}
\end{table}
\noindent In the matrix shown in \autoref{tab: dyadic_clust} above—full dyadic clustering—we can now see that the only observations assumed to be independent are those for which the dyad does not share any members with other observations. 

More specifically, our approach to DCRSEs follows \cite{aronow2015cluster}, which, in practice, allows for the DCR variance estimator to be decomposed entirely using robust variance estimators that are readily computed using popular statistical software packages.\footnote{This “multi-way decomposition” approach was first introduced by \cite{cameron2011robust}, and is based on the realization that each dyad member “is the basis of its own cluster that intersects with other units’ clusters,”  and that each of those dyad-member-specific clusters can be adjusted for using conventional cluster-robust variance estimators \citep{aronow2015cluster}. DCRSEs are mathematically equivalent to multi-way clustering on (undirected) dyad-member-specific clusters, though the large number of such multi-way clusters implies important analytic and computational distinctions.} The DCR variance estimator in this decomposed form is

\begin{align*}
\hat{V}_r = \sum_{i=1}^N \hat{V}_{c,i} - \hat{V}_D - (N-2)\hat{V}_0 
\end{align*}

\noindent
where $\hat{V}_r$ is the estimated DCR variance-covariance matrix for longitudinal data; $\hat{V}_{c,i}$ is the estimated dyad-member-\textit{i}-specific clustered variance-covariance matrix; $\hat{V}_D$ is the estimated repeated-dyad clustered variance-covariance matrix; and $\hat{V}_0$ is the estimated heteroskedasticity-consistent variance-covariance matrix. Taking the square root of the diagonal of the DCR variance-covariance matrix yields DCRSEs for all model parameters.\footnote{To account for rare instances where negative parameter estimate variances arise due to non-positive semi-definite variance-covariance matrices, an “eigendecomposition of the estimated variance matrix … [that] converts any negative eigenvalue(s) to zero” and then reassembles the matrix to “force” positive semi-definiteness is performed, in line with suggestions from \cite{cameron2011robust} and \cite{cameron2014robust}.}

Limit theorems for dyadic data \citep{tabord2019inference} establish that DCRSEs may be used to form asymptotically valid confidence intervals and $p$-values under a normal approximation. However, DCRSEs will tend to have more sampling variability than will conventional estimators that impose more structure (e.g., standard CRSEs), meaning that they may be unreliable for inference in small samples. Although simple corrections exist \citep[c.f.,][]{berge2021on,cameron2011robust}, theory and further refinements for small samples \citep[e.g.,][]{imbens2016robust, pustejovsky2018small} remain topics of ongoing inquiry for multiway clustering problems, including the dyadic clustering problem.

Note that the above approach and its implementation only corrects for interdependence between shared countries. DCRSEs are \textit{not} sufficient---although still improve over the naive approach or clustering by repeated dyad approach---when there are interdependencies throughout the entire system (i.e., across non-incident dyads). Two common examples of such systemic interdependencies in IR are alliance formation and multilateral trade deals. In deciding to form an alliance, friendly countries $i$ and $j$ may be influenced by a previous alliance formed by countries $i$ and $k$ (i.e., the $ij$ alliance is more attractive now that $ik$ are also allied). However, the $ij$ alliance could also be influenced by an alliance between unfriendly countries $a$ and $b$. DCRSEs do not capture the $ij$-$ab$ interdependence as there are no common dyad members.  Likewise, a multilateral trade deal may also be influenced by multiple pairs of relationships that do not necessarily share members.\footnote{See \citet{poast2016dyads} for a more thorough discussion.} \citet{aronow2018confidence}  developed conservative estimators for the variance of least squares estimators in such cases where there is further dependence---and in such cases we expect the variance of least squares estimates to be larger than those that only take into account dyadic clustering. There is therefore still a need to understand the data generating process and underlying theory of an empirical setting prior choosing an estimation strategy.\footnote{Political scientists also employ network analysis approaches that model dependence beyond incident dyads. For example, \citet{duque2018recognizing} uses a (temporal) exponential random graph model (ERGM) to examine the determinants of embassy formation; \citet{schoeneman2022complex} use an ERGM approach to investigate FDI networks; \citet{cho2021climate} uses ERGM to examine legislative co-sponsorship networks; \citet{kinne2020guns} use a stochastic actor oriented model (SOAM) to study how defense and economic cooperation are related; and \citet{dorff2020networks} use latent space methods (additive and multiplicative effects, or AME) to predict conflict between groups. Some of these approaches pertain to network formation models---not models of traits conditional on networks---and therefore do not relate to error terms in the context of dyadic networks. Other approaches do pertain to modeling traits and outcomes conditional on networks, and researchers may have interest in applying or adapting these methods to their own use cases when appropriate.}

\section{Reanalyzing Previous Studies}

To study the consequences of failing to account for dyadic clustering \textit{in practice}, we reanalyze recent, prominent studies from the international relations literature by applying DCRSEs to estimates in empirical contexts where DCRSEs are uniquely suited to handle dyadic clustering.\footnote{See \citet{carlson2023data} for a complete replication package.}

Specifically, we reanalyze all empirical articles published in \textit{International Organization} over a period of six years (from January 2014 through January 2020) that feature dyadic data. Studies were discovered by performing a Google Scholar search of all publications mentioning any form of the word ``dyad'' in this period. Specifically, a search query specified the publication name ``International Organization" and keywords ``dyadic OR dyad OR dyads." This process returned 70 candidate studies for reanalysis.

Each study was then assessed to determine its susceptibility to dyadic clustering. Studies were excluded from reanalysis for three primary reasons: (1) the study did not actually analyze dyadic data; (2) dyadic observations primarily featured nested relationships between dyad members, for which single or multi-way clustering of standard errors is sufficient\footnote{For example, consider model parameter estimates derived from dyadic data on civil wars consisting of rebel-state dyads. Because rebel groups are commonly nested within states, correlations across observations in the data can be accounted for using only state clustered standard errors.}; and (3) the dyadic observations featured a common dyad member across all observations, e.g., a nominally dyadic dataset consisting entirely of U.S. bilateral trade flows. There were 22 studies not excluded by these criteria (see \autoref{tab: paper_res} for a list of included studies). For each of these studies, replication data was either publicly available or provided by the authors.

For each eligible study, models that featured a key explanatory variable (KEV)\citep{lall2016multiple} fit to dyadic data were identified and reanalyzed. For this reanalysis, KEVs are defined as independent variables whose parameter estimates are directly referenced in the study, or otherwise clearly pertain to the study’s stated hypotheses. Control variables are not considered to be KEVs, even if discussed or directly referenced in the study. Specifically, a model was selected for reanalysis if: (1) the model was dyadic; (2) a KEV appeared in the model; and (3) the model was not relegated to an analysis explicitly denoted as a robustness check, sensitivity analysis, or supplementary analysis, unless one of these analyses was the only dyadic analysis in the study.
	
The final analytic sample consisted of 691 KEVs across 174 models from 22 studies.\footnote{The large number of KEVs are primarily driven by two outlier studies---\citet{bermeo2017aid} and \citet{goemans2017politics}---which have 173 and 170 KEVs, respectively. We control for these outliers by weighting our analyses by the inverse of the total number of KEVs for a given study appearing in the analytic sample.} While many studies clustered standard errors on repeated dyads, none utilized a non-parametric DCR variance estimation strategy in conducting primary analyses. Only 24 models across three studies did not use any sort of robust or clustered standard error.\footnote{See \autoref{tab: SERnoCRSE} for a summary of the results of DCR estimation for these models when repeated dyad cluster robust standard errors are added in.} Three studies employ at least one model that has fixed effects for both members of the dyad under consideration.\footnote{\autoref{tab: fe_table} shows the results of applying DCRSEs to the KEVs in these three papers.}  Of the 22 studies, 20 perform analysis on state (country) dyads and two perform analysis on international organization (IO) dyads.

All models were replicated and then re-estimated using the previously discussed DCR variance estimator formulated in \citet{aronow2015cluster} and implemented using an original suite of functions and commands for \textsf{R} and Stata, respectively. To ensure the comparability of results, the replications and reanalyses of selected models were conducted using the statistical software of origin. For the purposes of our reanalysis, dyads are assumed to be undirected, as modeling dyadic clustering based on directed dyads would require a stronger assumption about independence across observations.\footnote{Note that in the event that the dyadic information is indeed directional, we simply recover the directed dependence structure.} Also for the purposes of our primary reanalysis, there are no small-sample corrections made to our standard error estimates. Finite-sample corrections would inflate our standard error estimates to account for increased sampling variability, and thus might paint an overly pessimistic picture of the original literature.\footnote{In \autoref{tab: mega_agg_dof}, we repeat our reanalysis with the small-sample correction based on guidance from, e.g., \citet{berge2021on} and \citet{cameron2011robust}, and discover no substantive differences in findings with respect to our primary reanalysis.}

\subsection{Results}

To quantify the impact of neglecting dyadic clustering in prior empirical IR findings, we compare DCR re-estimated standard errors with non-DCR standard errors for all KEVs in all models. We compute a standard error ratio (SER) for all KEVs, which is the DCRSE divided by the standard error produced using the original variance estimation strategy.\footnote{For example, if the original model clustered standard errors on repeated dyads, the SER is a ratio of DCR standard error to repeated dyad clustered standard error. The SER can therefore be interpreted as the inflation (or deflation) of the original standard error introduced by inter-dyad dependencies.} We also examine the precision of reanalyzed KEVs, with special attention paid to estimates that lose statistical significance at a conventional level (i.e., 5\%).

\begin{figure}[H]
\begin{centering}
\adjincludegraphics[width=11cm,trim={0 1cm 0 1cm},clip]{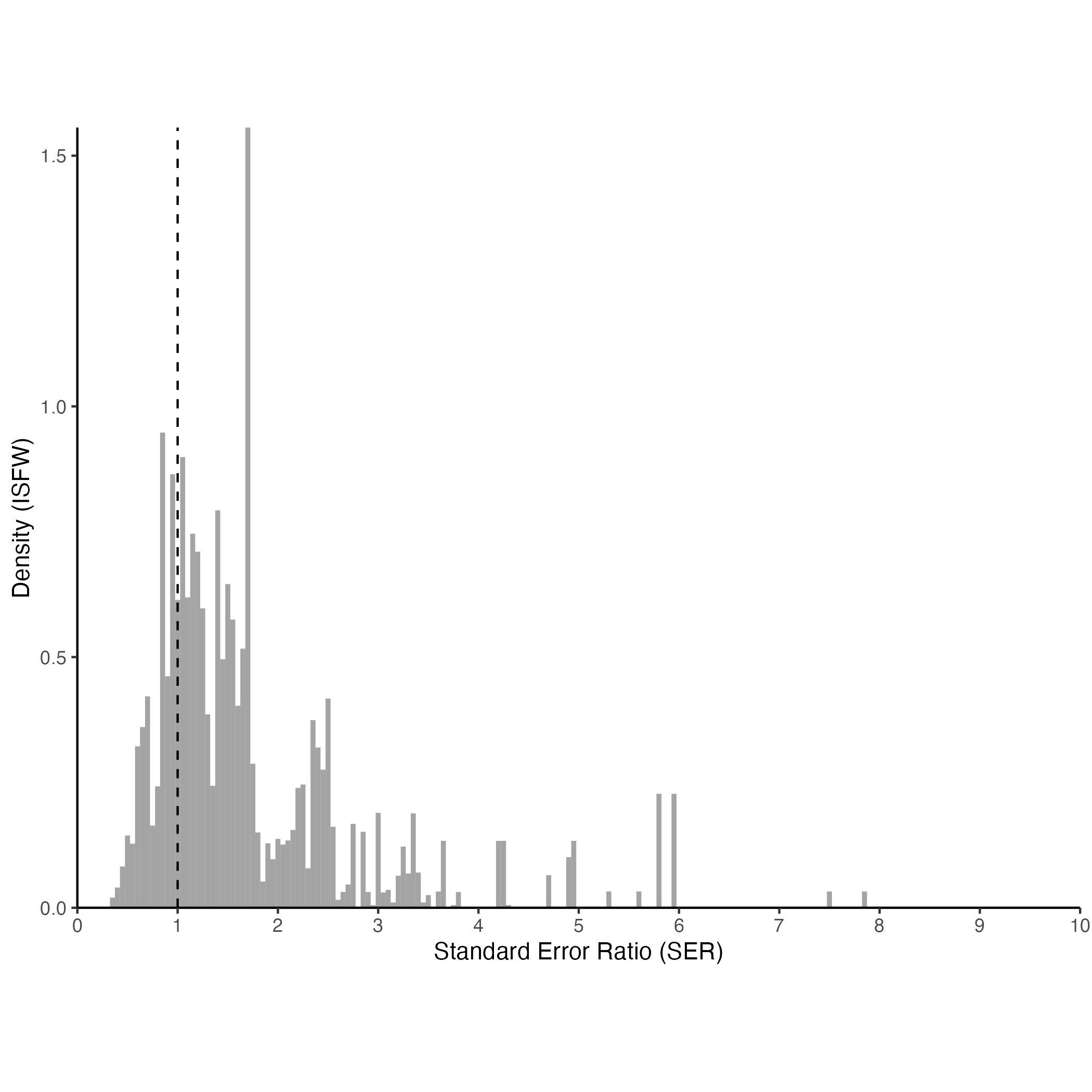}
\caption{Histogram of Key Explanatory Variable Standard Error Ratios}
\label{fig: ser_histogram}
\end{centering}
{\small Note: Key explanatory variables are independent variables whose parameter estimates are directly referenced in the study, or otherwise clearly pertain to the study’s stated hypotheses. Standard error ratios are the dyadic clustering robust standard errors divided by the standard error produced using the original variance estimation strategy.}
\end{figure}

The primary results of the reanalysis are presented as aggregated by year and subfield, as well as overall, in \autoref{tab: mega_agg}. The empirical distribution of SERs is presented in the histogram in \autoref{fig: ser_histogram}, and a breakdown of average SERs by study analyzed can be found in \autoref{tab: paper_res}. The inverse “study frequency” weighted (ISFW) average of SERs across all KEVs from all studies is 1.74, and the ISFW proportion of KEVs that go from significant to insignificant at the 5\% level across all studies is 0.22, where “study frequency” is the total number of KEVs for a given study appearing in the analytic sample.\footnote{Inverse ``study frequency" weighting is used to ensure that multi-study averages of SERs or multi-study proportions of significant results are not dominated by studies that contribute more KEVs to the analytic sample. ISFW is equivalent to an average of averages across studies.} Due to the sampling variability of the standard error estimators, we see a small but not zero proportion (0.05) of KEV estimates change from statistical insignificance to significance at the 5\% level.

We also examine which subfields suffer most from dyadic clustering. \autoref{tab: mega_agg} shows that no subfield is immune to standard error inflation. All subfields in our sample have an average SER indicating standard error inflation of more than 50\%, and all subfields see 19\% or more of all of their estimates change from statistically significant to insignificant at the 5\% level. The ``IOs and International Law" average SER indicates that DCRSEs are on average more than twice the size of originally reported standard errors, and that 34\% of estimates become insignificant with DCRSEs. 

\begin{table}[H]
\caption{Primary Results, Various Levels of Aggregation}
\label{tab: mega_agg}
\centering
\scalebox{0.8}{
\begin{threeparttable}
\begin{tabular}{llrrrrrrr}
\hline
Aggregation & Category & Studies & KEVs & SER\tnote{a} & Sig.$\rightarrow$Sig.\tnote{b} & Sig.$\rightarrow$Insig.\tnote{b} & Insig.$\rightarrow$Sig.\tnote{b} & Insig.$\rightarrow$Insig.\tnote{b} \\
\hline
Year & 2014 &   4 &  58 & 1.94 & 0.22 & 0.50 & 0.03 & 0.26 \\ 
& 2015 &   5 & 122 & 1.48 & 0.62 & 0.11 & 0.03 & 0.24 \\ 
& 2016 &   5 &  97 & 2.26 & 0.53 & 0.26 & 0.00 & 0.22 \\ 
& 2017 &   3 & 349 & 1.96 & 0.56 & 0.15 & 0.03 & 0.26 \\ 
& 2018 &   2 &  24 & 1.36 & 0.61 & 0.11 & 0.06 & 0.22 \\ 
& 2019 &   1 &  15 & 1.21 & 0.20 & 0.00 & 0.00 & 0.80 \\ 
& 2020 &   2 &  26 & 1.05 & 0.17 & 0.19 & 0.28 & 0.37 \\  
 \hline
Subfield & IOs/Int'l Law &  5 &  91 & 2.19 & 0.32 & 0.34 & 0.04 & 0.29 \\ 
  & IPE &   6 & 270 & 1.79 & 0.41 & 0.19 & 0.09 & 0.31 \\ 
  & Security & 11 & 330 & 1.51 & 0.54 & 0.19 & 0.02 & 0.25 \\ 
  \hline
All & All &  22 & 691 & 1.74 & 0.46 & 0.22 & 0.05 & 0.28 \\  
\hline
\end{tabular}
\begin{tablenotes}
\item[a] ``SER'' denotes an inverse “study frequency” weighted (ISFW) average of standard error ratios for a given level of aggregation. 
\item[b] ``Sig.$\rightarrow$Sig.," ``Sig.$\rightarrow$Insig.," ``Insig.$\rightarrow$Sig.," and ``Insig.$\rightarrow$Insig." denote ISFW proportions of $p$-values that change significance levels in these respective ways for a given level of aggregation.
\end{tablenotes}
\end{threeparttable}}
\end{table}

\begin{figure}[H]
\centering
\adjincludegraphics[width=11cm,trim={0 3cm 0 3cm},clip]{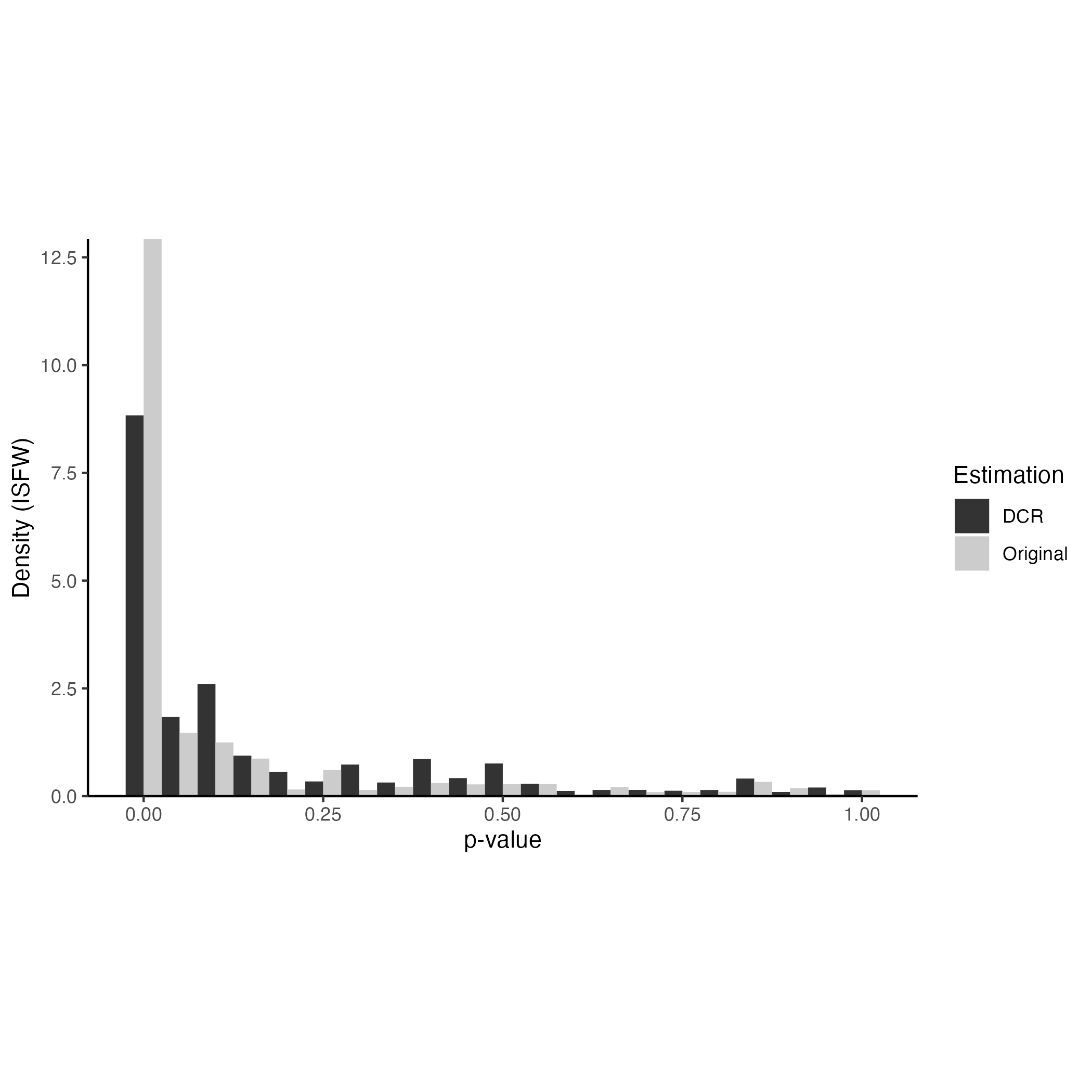}
\caption{Histogram of $p$-values before and after reanalysis using dyadic clustering robust standard errors}
\label{fig: pval_histogram}
\end{figure}

\begin{figure}[H]
\begin{centering}
\includegraphics[width=0.70\textwidth]{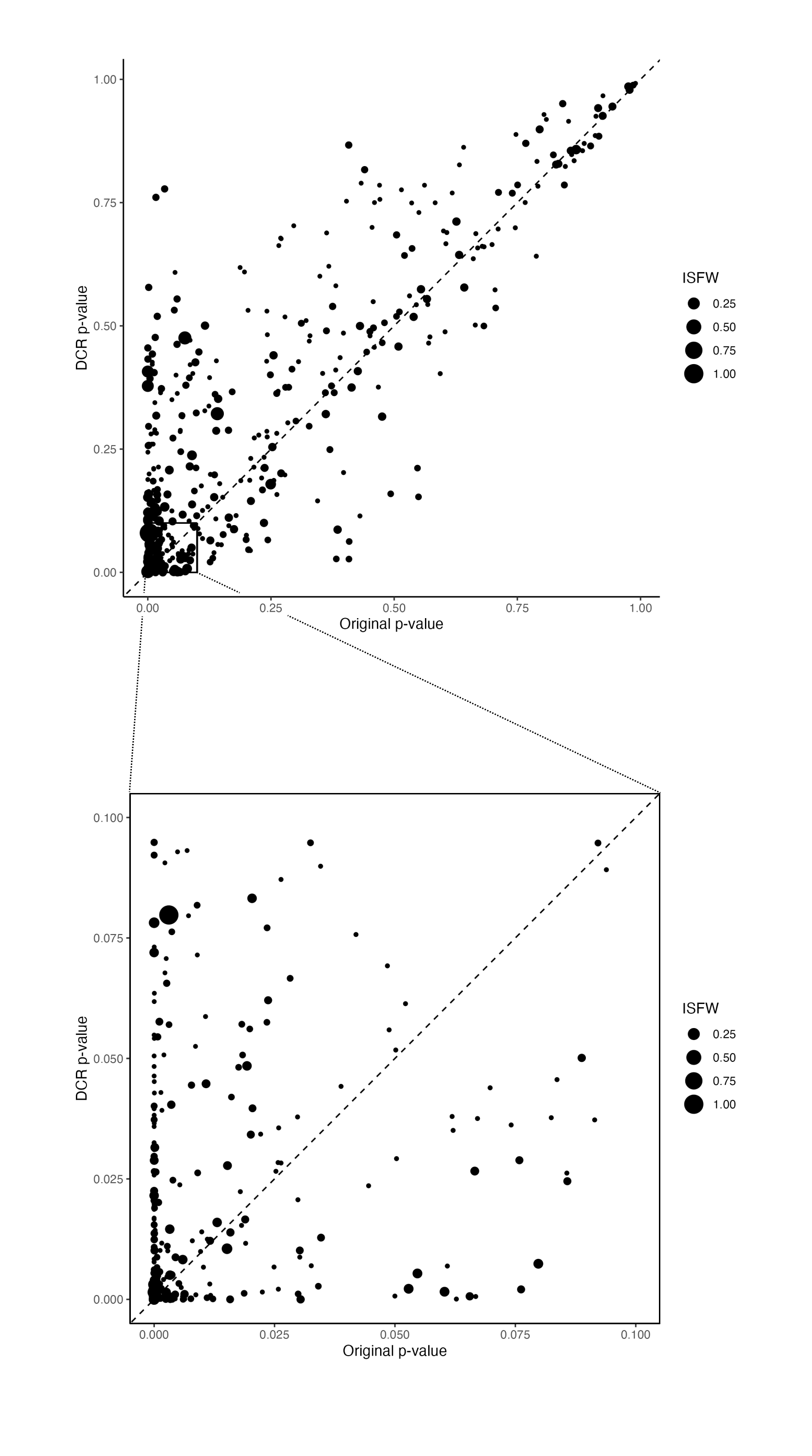}
\caption{Scatter plots of $p$-values before and after reanalysis using dyadic clustering robust standard errors}
\label{fig: scatter_pval}
\end{centering}
{\small Note: Top panel depicts all p-values in the reanalysis. Bottom panel depicts p-values below 0.1. Area of each plotted data point is proportional to its inverse study frequency weight (ISFW).}
\end{figure}

For individual studies, the average SER ranges from 0.90 to 4.16, as seen in \autoref{tab: paper_res}. On average, KEVs from 18 of 22 studies are less precise after accounting for dyadic clustering. Only 7 of 22 studies do not lose statistical significance in any KEV estimates due to DCRSE re-estimation. Three studies see half or more of their KEVs become insignificant  at the 5\% level upon reanalysis. \autoref{fig: pval_histogram} shows how the empirical distribution of KEV $p$-values from all reanalyzed studies shifts due to the application of DCRSEs. \autoref{fig: scatter_pval} visualizes how the precisions of individual KEV estimates change with the application of DCRSEs across all reanalyzed studies, with the area of each plotted data point being proportional to its inverse study frequency weight.

%\clearpage

\section{Conclusion}

Though the need to account for the complex dependencies in dyadic data has been noted by previous researchers, these recommendations have not been commonly applied in practice. We investigated the consequences of failing to account for dyadic clustering  in previous empirical research by reanalyzing all quantitative dyadic analyses in \textit{International Organization} published in a six year window, thereby revealing a lower bound on the severity of the consequences of inadequate clustering practices in previous quantitative dyadic research.

We find that the standard errors associated with key explanatory variables in reanalyzed studies are approximately half of what they would have been if calculated as DCRSEs, but that two-thirds of statistically significant KEVs remain statistically significant when using DCRSEs. Failure to compute DCRSEs therefore does not appear to have led to systematically false substantive conclusions in recent empirical IR literature, but can lead to a systematically large overestimation of the precision of estimates. In short, dyadic clustering matters, yet is not so severe as to make statistical inference using dyadic data infeasible. 

However, solely accounting for dyadic clustering may not go far enough. For any of the studies we reanalyze, dependencies may exist in the data that extend across non-incident dyads, e.g., due to network effects, in which case the problem may be even more severe than what our reanalysis suggests. That said, DCRSEs may be of particular interest to researchers because dyadic clustering—the clustering structure associated with dependence across dyads that share a member—is a feature of many dyadic datasets of interest to social scientists. Accordingly, we offer software in \textsf{R} and Stata to facilitate future analyses robust to dyadic clustering. These open source packages implement DCRSEs for all the models in the reanalysis sample (and more), and mirror syntax familiar to users of \textsf{R} and Stata.

\section*{Acknowledgements}

We would like to thank Austin Jang for excellent and extensive research assistance, as well as Jonathon Baron, Laurent Bergé, Forrest Crawford, Joshua Kalla, Winston Lin, Cleo O’Brien-Udry, Paul Goldsmith-Pinkham, Cyrus Samii, Beth Tipton, and three anonymous reviewers for helpful comments and conversations.

\section*{Data Availability Statement}

Replication code for this article can be accessed via Dataverse \citep{carlson2023data}. The statistical programming suite for DCR estimation is also available. To access the source code for the \verb+dcr+ command for Stata (version 15 or higher), clone its GitHub repository: 
\begin{verbatim}
git clone git@github.com:jscarlson/stata-dcr.git \end{verbatim} 
\noindent
\noindent
To access the \verb+dcr+ package for \textsf{R}, run: \begin{verbatim}
devtools::install_github("jscarlson/dcr") \end{verbatim}

%\section*{Funding}
%This research was supported in part by ABC.

\section*{Supplementary Material}

For supplementary material accompanying this paper, please visit the Political Analysis Harvard Dataverse \citep{carlson2023data}.

%%%%%% Begin Appendices %%%%%%%

\newpage

\printbibliography

\appendix
\setcounter{secnumdepth}{1}
\setcounter{table}{0}
\setcounter{figure}{0}
\renewcommand\thetable{\Alph{section}.\arabic{table}}
\renewcommand\thefigure{\Alph{section}.\arabic{figure}}

\newpage

\begin{centering}
\LARGE
\textbf{\textit{Supplementary Appendix:}} \\
 \vspace{0.5cm}
\textbf{Dyadic Clustering in International Relations} \\

\vspace{0.5cm}

\large
Jacob Carlson, Trevor Incerti, and P.M. Aronow
 
\end{centering}

\vspace{2cm}

\section{Appendix} \label{Appendix}
\localtableofcontents
\pagenumbering{arabic}% resets `page` counter to 1
\renewcommand*{\thepage}{A\arabic{page}}

\vspace{2cm}

\subsection{Statistical Programming Suite}
\noindent The statistical programming suite for DCR estimation is currently available. To access the source code for the \verb+dcr+ command for Stata (version 15 or higher), clone its GitHub repository: 
\begin{verbatim}
git clone git@github.com:jscarlson/stata-dcr.git \end{verbatim} 
\noindent
\noindent
To access the \verb+dcr+ package for \textsf{R},  run: \begin{verbatim}
devtools::install_github("jscarlson/dcr") \end{verbatim}

%%%

\clearpage
\subsection{Supplementary Figures and Tables}

\begin{table}[ht]
\caption{Dyadic Data Summary Statistics, Study Level}
\label{tab: paper_summary}
\centering
\scalebox{0.8}{
\begin{threeparttable}
\begin{tabular}{lllll}
\hline
  Study & Observations & Data Start Year & Data End Year & Number of Unique Dyads \\ 
  \hline
 Arel-Bundock (2017) & 3971 & 2012 & 2012 & 2161 \\ 
   Bermeo (2017) & 20275 & 1973 & 1988 & 1614 \\ 
   Bermeo and Leblang (2015) & 33116 & 1993 & 2008 & 3129 \\ 
   Carter and Poast (2020) & 40002 & 1948 & 2011 & 560 \\ 
    Colgan (2014) & 213454 & 1949 & 1999 & 37310 \\ 
    Colgan and Weeks (2015) & 804434 & 1951 & 2000 & 28410 \\ 
    Copelovitch and Putnam (2014) & 101 & 1926 & 1986 & 92 \\ 
    Dietrich (2016) & 10605 & 2006 & 2011 & 2417 \\ 
    Efrat and Newman (2016) & 13818 & 1996 & 2012 & 2574 \\ 
    Fang, Johnson, and Leeds (2014) & 585467 & 1816 & 2000 & 21133 \\ 
    Goemans and Schultz (2017) & 153370 & 1958 & 1998 & 102 \\ 
    Horowitz and Stam (2014) & 113992 & 1875 & 2000 & 27079 \\ 
    Kinne (2018) & 253296 & 1981 & 2010 & 10909 \\ 
    McDonald (2015) & 525114 & 1817 & 2000 & 13938 \\ 
    Powell (2015) & 433 & 1948 & 2006 & 40 \\ 
    Pratt (2018) & 1177 & 2002 & 2012 & 108 \\ 
    Renshon (2016) & 1237676 & 1818 & 2001 & 39389 \\ 
    Schneider and Tobin (2020) & 4760 & 1976 & 2009 & 756 \\ 
    Shelef (2016) & 344 & 1945 & 1995 & 155 \\ 
    Sommerer and Tallberg (2019) & 37989 & 1971 & 2009 & 2080 \\ 
    Weisiger (2016) & 36322 & 1823 & 2001 & 81 \\ 
    Weisiger and Yarhi-Milo (2015) & 1187663 & 1816 & 2000 & 33255 \\ 
    \hline
\end{tabular}
\begin{tablenotes}
\small
 \item Note: summary statistics for each study computed using data from a representative (usually the first) reanalyzed model, as analytic samples and data often vary by model.
 \end{tablenotes}
\end{threeparttable}}
\end{table}

\begin{landscape}

\begin{table}
\caption{Primary Results, Study Level}
\label{tab: paper_res}
\centering
\scalebox{0.6}{
\begin{tabular}{lllllllllll}
\hline
  Study & Subfield & Secondary Subfield & Primary Relationship Tested & KEVs & Sig.$\rightarrow$Sig. & Sig.$\rightarrow$Insig. & Insig.$\rightarrow$Sig. & Insig.$\rightarrow$Insig. & Avg. SER \\ 
  \hline
Arel-Bundock (2017) & IPE & International finance & Treaty shopping and domestic tax rates & 6 & 0.833 & 0.167 & 0 & 0 & 2.739 \\ 
Bermeo (2017) & IPE & Foreign aid & (Donor spillovers and) foreign aid & 173 & 0.445 & 0.249 & 0 & 0.306 & 2.187 \\ 
Bermeo and Leblang (2015) & IPE & Foreign aid & (Immigration and) foreign aid & 29 & 0.655 & 0.207 & 0 & 0.138 & 2.2 \\ 
Carter and Poast (2020) & IPE & Trade & Physical barriers and legal trade flows & 8 & 0 & 0.375 & 0.5 & 0.125 & 1.099 \\ 
Colgan (2014) & IOs/International Law & IO efficacy & IO membership and member behavior & 9 & 0.111 & 0.889 & 0 & 0 & 3.269 \\ 
Colgan and Weeks (2015) & Security & Conflict & (Revolution and) likelihood of conflict & 24 & 1 & 0 & 0 & 0 & 1.526 \\ 
Copelovitch and Putnam (2014) & IOs/International Law & IO efficacy & Rational design theory/international agreement formation & 45 & 0.422 & 0.111 & 0.111 & 0.356 & 0.898 \\ 
Dietrich (2016) & IPE & Foreign aid & Variety of capitalism and foreign aid delivery method & 36 & 0.194 & 0.139 & 0 & 0.667 & 1.511 \\ 
Efrat and Newman (2016) & IOs/International Law & IO efficacy & Likelihood of defering to another country's laws & 4 & 0.5 & 0.5 & 0 & 0 & 4.158 \\ 
Fang, Johnson, and Leeds (2014) & Security & Conflict & (Alliances and) likelihood of conflict & 1 & 0 & 1 & 0 & 0 & 1.691 \\ 
Goemans and Schultz (2017) & Security & Conflict & Ethnic politics and territorial claims & 170 & 0.394 & 0.041 & 0.088 & 0.476 & 0.955 \\ 
Horowitz and Stam (2014) & Security & Conflict & (Leader military experience) and likelihood of conflict & 3 & 0.333 & 0 & 0 & 0.667 & 1.895 \\ 
Kinne (2018) & Security & Alliance formation & Determinants of defense cooperation agreements & 6 & 0.833 & 0 & 0 & 0.167 & 1.301 \\ 
McDonald (2015) & Security & Conflict & (Democracy and) likelihood of conflict & 41 & 0.488 & 0.195 & 0 & 0.317 & 1.432 \\ 
Powell (2015) & Security & Conflict & (Islamic law and) likelihood of conflict & 16 & 0.188 & 0 & 0.062 & 0.75 & 0.908 \\ 
Pratt (2018) & IOs/International Law & IO efficacy & Likelihood of defering to another IO & 18 & 0.389 & 0.222 & 0.111 & 0.278 & 1.426 \\ 
Renshon (2016) & Security & Conflict & (International status and) likelihood of conflict & 20 & 0.9 & 0.05 & 0 & 0.05 & 1.291 \\ 
Schneider and Tobin (2020) & IPE & International finance & Domestic spillovers and international bailout & 18 & 0.333 & 0 & 0.056 & 0.611 & 0.999 \\ 
Shelef (2016) & Security & Conflict & (Type of territorial dispute and) likelihood of conflict & 9 & 0.889 & 0 & 0 & 0.111 & 1.071 \\ 
Sommerer and Tallberg (2019) & IOs/International Law & IO design & IO connectivity and convergence of governance rules & 15 & 0.2 & 0 & 0 & 0.8 & 1.212 \\ 
Weisiger (2016) & Security & Conflict & Determinants of conflict duration & 28 & 0.143 & 0.607 & 0 & 0.25 & 3.268 \\ 
Weisiger and Yarhi-Milo (2015) & Security & Conflict & (Past actions and) likelihood of conflict & 12 & 0.75 & 0.167 & 0.083 & 0 & 1.311 \\ 
 \hline
Average-of-Averages (= ISFW) &  &  &  &  & 0.455 & 0.224 & 0.046 & 0.276 & 1.743 \\ 
 \hline
 \end{tabular}}
\end{table}
\end{landscape}

%%%

\begin{table}[ht]
\caption{Reanalysis for Models without (C)RSEs}
\label{tab: SERnoCRSE} 
\centering
\resizebox{\linewidth}{!}{
\begin{threeparttable}
\begin{tabular}{lrrrrrr}
  \hline
  Study & KEVs & Sig.$\rightarrow$Sig. & Sig.$\rightarrow$Insig. & Insig.$\rightarrow$Sig. & Insig.$\rightarrow$Insig. & Avg. SER \\ 
  \hline
Kinne (2018) &   4 & 0.75 & 0.00 & 0.00 & 0.25 & 1.03 \\ 
Copelovitch and Putnam (2014) &  45 & 0.42 & 0.11 & 0.11 & 0.36 & 0.90 \\ 
Fang, Johnson, and Leeds (2014) &   1 & 0.00 & 1.00 & 0.00 & 0.00 & 1.69 \\ 
   \hline
\end{tabular}
\begin{tablenotes}
\small
 \item Note: repeated dyad CRSEs computed as comparison to DCRSEs.
 \end{tablenotes}
\end{threeparttable}}
\end{table}

\begin{table}[ht]
\caption{Reanalysis for Models with Fixed Effects for both Dyad Members}
\label{tab: fe_table}
\centering
\resizebox{\linewidth}{!}{
\begin{tabular}{lrrrrrr}
  \hline
 Study & KEVs & Sig.$\rightarrow$Sig. & Sig.$\rightarrow$Insig. & Insig.$\rightarrow$Sig. & Insig.$\rightarrow$Insig. & Avg. SER \\ 
  \hline
Carter and Poast (2020) &   8 & 0.00 & 0.38 & 0.50 & 0.12 & 1.10 \\ 
Shelef (2016) &   9 & 0.89 & 0.00 & 0.00 & 0.11 & 1.07 \\ 
Bermeo and Leblang (2015) &   1 & 1.00 & 0.00 & 0.00 & 0.00 & 2.13 \\ 
   \hline
\end{tabular}}
\end{table}

\clearpage

%%%

\begin{table}[ht]
\caption{Small-Sample Corrected Results, Various Levels of Aggregation}
\label{tab: mega_agg_dof}
\centering
\resizebox{\linewidth}{!}{
\begin{threeparttable}
\begin{tabular}{llrrrrrrr}
\hline
Aggregation & Category & Studies & KEVs & SER\tnote{a} & Sig.$\rightarrow$Sig.\tnote{b} & Sig.$\rightarrow$Insig.\tnote{b} & Insig.$\rightarrow$Sig.\tnote{b} & Insig.$\rightarrow$Insig.\tnote{b} \\
\hline
Year & 2014 &   4 &  58 & 1.94 & 0.21 & 0.51 & 0.02 & 0.26 \\ 
& 2015 &   5 & 122 & 1.48 & 0.62 & 0.11 & 0.03 & 0.24 \\ 
& 2016 &   5 &  97 & 2.27 & 0.50 & 0.28 & 0.00 & 0.22 \\ 
& 2017 &   3 & 349 & 1.97 & 0.55 & 0.15 & 0.03 & 0.27 \\ 
& 2018 &   2 &  24 & 1.38 & 0.58 & 0.14 & 0.06 & 0.22 \\ 
& 2019 &   1 &  15 & 1.23 & 0.20 & 0.00 & 0.00 & 0.80 \\ 
& 2020 &   2 &  26 & 1.01 & 0.11 & 0.19 & 0.33 & 0.37 \\ 
 \hline
Subfield & IOs/International Law &   5 &  91 & 2.21 & 0.31 & 0.36 & 0.04 & 0.29 \\ 
& IPE &   6 & 270 & 1.78 & 0.39 & 0.19 & 0.11 & 0.31 \\ 
& Security &  11 & 330 & 1.52 & 0.53 & 0.20 & 0.02 & 0.26 \\ 
  \hline
 All & All &  22 & 691 & 1.75 & 0.44 & 0.23 & 0.05 & 0.28  \\  
\hline
\end{tabular}
\begin{tablenotes}
\item Note: this reanalysis applies a small-sample correction to the final estimates of DCRSEs. This correction is equivalent to multiplying the final estimates of standard error by $\sqrt{N/(N-1)}$, where $N$ is the number of unique dyad members in the analytic sample. Correspondingly, when computing $p$-values, the test statistic is compared to a t-distribution with degrees-of-freedom = $N-1$. For the purposes of producing a true apples-to-apples comparison, original standard error estimates of reanalyzed studies are similarly treated with a small-sample correction that is equivalent to multiplying the final estimates of standard error by $\sqrt{N_c/(N_c-1)}$, where $N_c$ is the number of unique clusters in the analytic sample; the test statistic is likewise compared to a t-distribution with degrees-of-freedom = $N_c-1$.
\item[a] ``SER'' denotes an ISFW average of SERs for a given level of aggregation. 
\item[b] ``Sig.$\rightarrow$Sig.," ``Sig.$\rightarrow$Insig.," ``Insig.$\rightarrow$Sig.," and ``Insig.$\rightarrow$Insig." denote ISFW proportions of $p$-values that change significance levels in these respective ways for a given level of aggregation.
\end{tablenotes}
\end{threeparttable}}
\end{table}

\end{document}